\documentclass[11pt]{article}
\usepackage{moriond,epsfig}

\def\be{\begin{equation}}
\def\ee{\end{equation}}
\def\bea{\begin{eqnarray}}
\def\eea{\end{eqnarray}}

\def\mutoe{$\mu^+ \to e^+ \gamma $~}

\def\rdm{$\mu^+ \to e^+ \nu_{e} \bar{\nu}_{\mu} \gamma$~}
\def\mich{$\mu^+ \to e^+ \nu_{e} \bar{\nu}_{\mu}$~}
\begin{document}
\vspace*{4cm}
\title{FIRST RESULT FROM THE MEG EXPERIMENT}

\author{ ELISABETTA BARACCHINI  \emph{on behalf of the MEG Collaboration}}

\address{University of California, Dept. of Physics and Astronomy, 4129 Frederick Reines Hall, \\
Irvine, CA 92697-4575 USA}

\maketitle\abstracts{We present the first results from the MEG experiment for 
the search of the Lepton Flavour Violating decay \mutoe. LFV decays are forbidden 
in the SM and highly suppressed in any minimal SM extension with tiny neutrino masses.
 On the other hand, several SUSY, GUT and ED models beyond the SM predict the 
 \mutoe rate at a level experimentally accessible. Hence, the MEG experiment 
 will be able either to provide an incontrovertible evidence of physics beyond the SM or to 
 significantly constrain the parameter space of SM extensions.}

\section{Introduction}
We report here the first result from the MEG experiment for the search of the 
Lepton Flavour Violating (LFV) process \mutoe at the 590 MeV proton ring at the 
Paul Scherrer Institute (PSI) in Switzerland. 

 Lepton Flavour Conservation is an accidental
symmetry of the Standard Model (SM), that is not required by the gauge structure of 
the theory, and is thus naturally violated in many SM extensions. LFV is actually already
been observed in the neutral sector, through neutrinos oscillations~\cite{Schwetz:2008er}. 
In minimal extension of the SM, LFV in the charged sector is mediated by massive neutrinos, giving prediction for 
the branching ratio ($BR$) beyond any experimental reach ($BR$(\mutoe) $\simeq m_{\mu}^4/m_{W}^4 \simeq 10^{-54}$).
On the other hand, heavy particles beyond the SM 
entering into virtual loops, can lead to huge enhancement of the rate, allowing
the measurement of the $BR$ to be  experimentally accessible ($BR$(\mutoe) $\simeq 10^{-12} - 10^{-14}$)
~\cite{Barbieri:1995tw,Hisano:1997tc,Marciano:2008zz}.
 At present, the most stringent upper limit (UL) on LFV in the charged sector comes from the 
muon system and has been set by the MEGA experiment as $BR$(\mutoe) $< 1.2 \times 10^{-11}(90 \% \rm{C.L.})$~\cite{Brooks:1999pu}.


The experimental signature of a \mutoe decay it is characterized by a very simple two-body
final state, where the photon and the positron are emitted coincident in time
and back to back in the muon rest frame, each with an energy
 equal to half the muon mass.
 
  There are two major backgrounds to this process:
 the radiative muon decay (RDM) \rdm and the accidental coincidence between
 an high energy positron from the Michel muon decay \mich and an high energy
 photon from RMD events, positrons annihilating in flight or bremsstrahlung. 
While the signal and RDM background are proportional to the muon rate, the
accidental background goes as the rate squared and thus dominates. It can
be shown~\cite{Kuno:1999jp} that the rate of accidental events is also linearly 
proportional to the positron energy resolution and the positron-photon time
difference resolution and goes quadratically with the photon energy and the
solid angle resolution. 

 Thus, in order to be able to reach an ${\cal O} (10^{-13})$
sensitivity, it is mandatory to have a continuos and intense muon beam able
to provide high statistic and a precision detector with excellent spatial, time
and energy resolutions.

\section{The MEG Detector}\label{sec:detector}
The MEG detector is schematically
composed of a solenoid spectrometer with low-mass drift chambers for the measurement of positron
energy and position, scintillators bars and fibers for the measurement of positron time
and a liquid Xenon calorimeter for the photon detection.

 A very high rate continuos beam ($\sim 3 \times10^7$ $\mu$/s) of surface muons
at 28 MeV/c from one of the world's most intense sources (the $\pi$E5 line
at PSI) is stopped in a thin polyethylene target surrounded by the MEG
spectrometer. A Wien filter that assures 7.5 $\sigma$ separations of muons
w.r.t. other particle that may pollute the beam and a superconducting transport
solenoid (BTS) to couple the beam to the detector magnetic field are employed 
in order to guarantee the most pure and intense beam possible. 

 Positron originating
from the target enter the COBRA (COstant Bending RAdius) superconducting
magnet, which provides a graded magnetic field ranging from 1.27 Tesla in the center
to 0.49 Tesla at each end. The field is designed such that positrons emitted with
the same momentum, follow trajectories with an almost constant projected bending
radius, independent of their emission angle. This allows a preferential acceptance of
higher momentum particles in the drift chambers, as well as sweeping away particles
more efficiently, compared to an uniform field, which is an essential feature in order to fight pile up.

 The drift chamber (DC) system is composed by 16 radially aligned modules spaced 
at $10.5^o$ intervals, forming and half-circle around the target. Each module
contains two staggered layers of anode wire planes, of nine drift cells each, and the layers
are separated and enclosed by 12.5 $\mu$m of Mylar cathode foils with a Vernier pattern
structure, for the determination of the $z$ coordinate. The chamber are filled with a 50:50
helium/ethane gas mixture, allowing for a total mass of $2.0 \times 10^{-3} X_{0}$ along
the positron trajectory. The goal resolutions of the drift chambers are around 200 $\mu$m
for X and Y single hit position and 400 $\mu$m for Z, and 200 KeV of resolution for the positron
momentum. 

 The positron time is measured by a scintillator timing counter arrays (TC), composed by two section
(upstream and downstream w.r.t. the target) placed at the end of the spectrometer. Each array consists
of 15 BC404 plastic scintillator bars, with 128 orthogonally placed BCF-20 scintillating fibers.
The bars are read by fine-mesh photomultiplier, while the fibers are viewed by avalanche photodiodes.
The TC is crucial for the positron time measurement and provides also impact point information
and direction, to be used in the trigger. The goal for the intrinsic time resolution of the TC is
about 50 ps. 

 The photon detector (XEC) is the largest liquid Xenon (LXe) calorimeter in the world ( about 900 L)
and covers a solid-angle acceptance of $\sim 10^{o}$. It exploits the scintillation light to measure
the energy of the photon and the position and time of the first interaction. The light is read by 846 
photomultipliers, which are mounted on the internal surface of the XEC. The uniformity, 
the fast response and the hight light yield ($\sim 75\%$ of NaI) of the LXe allow very good resolutions,
provided that the LXe is kept very pure: in fact, vaccum ultra-violet scintillation light is very easily
absorbed by water or oxygen even at sub-ppms levels. The xenon is therefore circulated in liquid
phase through a series of purification cartridges and in gas phase through a heated getter in order
to prevent contaminations. 

 The optical properties of the xenon, as well as of  PMTs gains and quantum efficiency
are constantly monitored by means of LEDs and point-like $^{241}$Am $\alpha$-sources deposited
on wires inside the LXe active volume. At the same time, energy and relative time calibrations are
provided by studying charge exchange (CEX) processes ($\pi^- p \to \pi^0 n \to \gamma \gamma n$)
where, selecting a definite opening angle, it is possible to obtain two mono-energetic lines for 
calibration, that allow to measure the energy scale and the uniformity. The Dalitz decays
($\pi^0 \to \gamma e^+ e^-$) in this process provide also a calibration for the detector time
synchronization. Additional calibrations using a Cockcroft-Walton (CW) accelerator of protons
against a $Li_{2} B_4 O_7$ target provide low energy photons from $^7 Li(p,\gamma)^8 Be$
for monitoring the LXe energy scale, while the two coincident $\gamma$'s from 
$^{11} B(p,\gamma)^{12} C$ are detected simultaneously by the XEC and the TC and allow
for the determination of the XEC-TC time offset. The goal resolution of the XEC are about 800 KeV
for the photon energy, 65 ps for photon timing and 2-4 mm for the photon conversion point.

\section{The 2008 Physics Run}
The data sample presented here has been collected between September and December 2008
and correspond to $\sim 9.5 \times 10^{13}$ muons stopping on the target. During the data-taking
the light yield of the XEC was continuously increasing due to the purification of the LXe, which was
performed in parallel. This was carefully monitored thanks to the several calibrations available
and properly taken into account in the determination of the energy scale in the final analysis.
At the same time, the trigger thresholds were accordingly adjusted in order to guarantee an
uniform efficiency through the whole run.
Moreover, an increasing number of drift chambers suffered frequent 
high-voltage trips resulting in a substantial reduction of the overall positron efficiency by a factor
of three over the whole period. This was due to and hardware problem which implied
a long exposure of the chamber to an helium atmosphere. The chambers were all rebuilt in
2009 and did not show anymore signs of this effect.

\subsection{Events Selection and Resolutions}\label{sec:res}
A \mutoe candidate event is characterized by five kinematic distributions: positron and photon
energies ($E_e$ and $E_{\gamma}$ respectively), relative time between the photon and the
positron ($t_{e\gamma}$) and opening angle between the two ($\theta_{e\gamma}$ and 
$\phi_{e\gamma}$). 

 The positron energy scale and resolution are evaluated by fitting the kinematic
edge of the measured Michel positron energy spectrum at 52.8 MeV with the convolution
of the theoretical Michel spectrum with the energy dependent detector efficiency (extracted
from data) and the response function for mono-energetic positron (extracted from Monte Carlo
(MC) simulation of the signal \mutoe decay). The latter is well described by the sum of a core
and two tails components, all of the three Gaussian. The resolutions extracted from data are
374 KeV, 1.06 MeV and 2.00 MeV in sigma for the core and the two tails component, corresponding
to fractions of 60$\%$, 33$\%$ and 7$\%$ respectively. 
The uncertainty on the fitted parameters
is dominated by the systematic of the fit and is determined varying selection and fitting criteria.

 The photon energy scale and resolution is extracted from the CEX data, where a small
correction is applied in order to take into account the different background conditions in the LXe
volume during the operation of the pion beam. The photon energy resolution 
is asymmetric with a low energy tail due to photons converting in front
of the LXe sensitive volume. The resolution function is thus actually dependent on the position
of the photon conversion, mainly on the depth inside the detector ($w$). The average resolution
on deep events ($w > $ 2 cm) is measured to be $\Delta E/E = (5.8 \pm 0.4) \%$ FWHM with
a right tail of $\sigma_R = (2.0 \pm 0.2) \%$, where the error takes into account the variation
over the acceptance. The photon energy scale is extracted from the measurement of the
17.67 MeV energy peak of the $^7 Li(p,\gamma)^8 Be$ reaction obtained with the CW protons
on the Li target. The scale is also confirmed by a fit to the photon energy spectrum taking into
account the expected spectra of RDM, positron annihilation in flight and photons pile-up, folded
with the shape determined from the CEX run. The difference between these two measurement
gives the energy scale systematic uncertainty, which is $< 0.4 \%$.

 The positron time is measured by the scintillator in the TC and corrected by the time-of-flight
of the positron from the target to the TC, as determined by the track length measured by the 
spectrometer. The photon time is determined from the rising of the waveforms in the XEC PMTs
and corrected by the line-of-flight that starts from the positron vertex at the target (as determined
by the spectrometer) and ends at the reconstructed conversion point in the XEC. 
The $t_{e\gamma}$ peak is fitted in the region
40 $< E_{\gamma}<$ 45 MeV and, corrected by a small $E_{\gamma}$-dependence observed 
in the CEX runs, the timing resolution for a signal event is estimated to be 
$\sigma_{t_{e\gamma}} = (148 \pm 17)$ ps. 

 The positron direction and decay vertex position are determined by projecting back to
the target the positron trajectory, as determined by the spectrometer. The photon direction is defined
by the line starting at the conversion point in the LXe and ending at the vertex of the candidate
companion positron. The full angular resolution is evaluated by combining the angular resolution 
and vertex resolution in the positron detector with the position resolution in the photon detector.
The positron angular resolution is determined by exploiting tracks that make two turns in the
spectrometer, where each turn is treated ad an independent track. The $\theta$ and $\phi$
~\footnote{In the coordinate system where the beam axis is the z-axis, $theta$ is defined as
the polar angle and $\phi$ as the azimuthal angle.} resolutions
are extracted separately from the difference of the two track segments at the point of closest approach
to the beam-axis and are $\sigma_{\theta}$ = 18 mrad and $\sigma_{\phi}$ = 10 mrad.
With the same technique, it is possible to evaluate the vertex decay position resolution to
be $\sim$ 3.2 mm and $\sim$ 4.5 mm in the vertical and horizontal directions on the target plane
respectively. These values are confirmed by an alternative estimation, which makes use of holes
placed in the target. The resolution on the photon conversion point is evaluated by MC simulation
and validated in dedicated CEX runs by placing different lead collimator in front of the LXe
volume. The average position resolution along the two orthogonal front-face sides of the LXe
and the depth $w$ are estimated to be $\sim$ 5 mm and $\sim$ 6 mm respectively.

\subsection{Analysis Technique}
A blind analysis technique has been adopted, where the events inside the signal region for the
photon energy and positron-photon time coincidence are not used for the analysis development
and only sideband data are employed  for background characterization.

\begin{figure}[t!]
\begin{center}
\includegraphics[width=6.3cm]{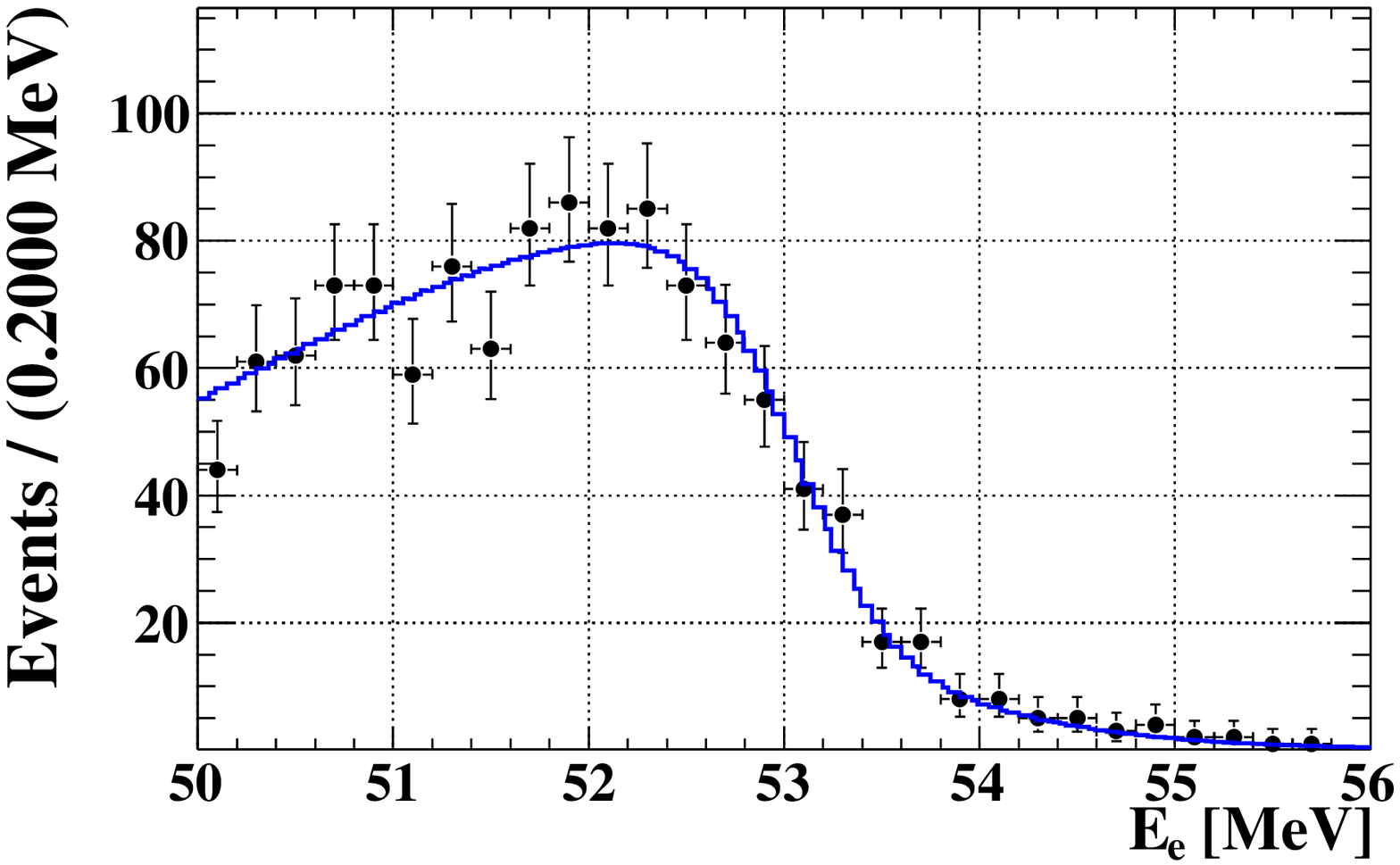}
\includegraphics[width=6.3cm]{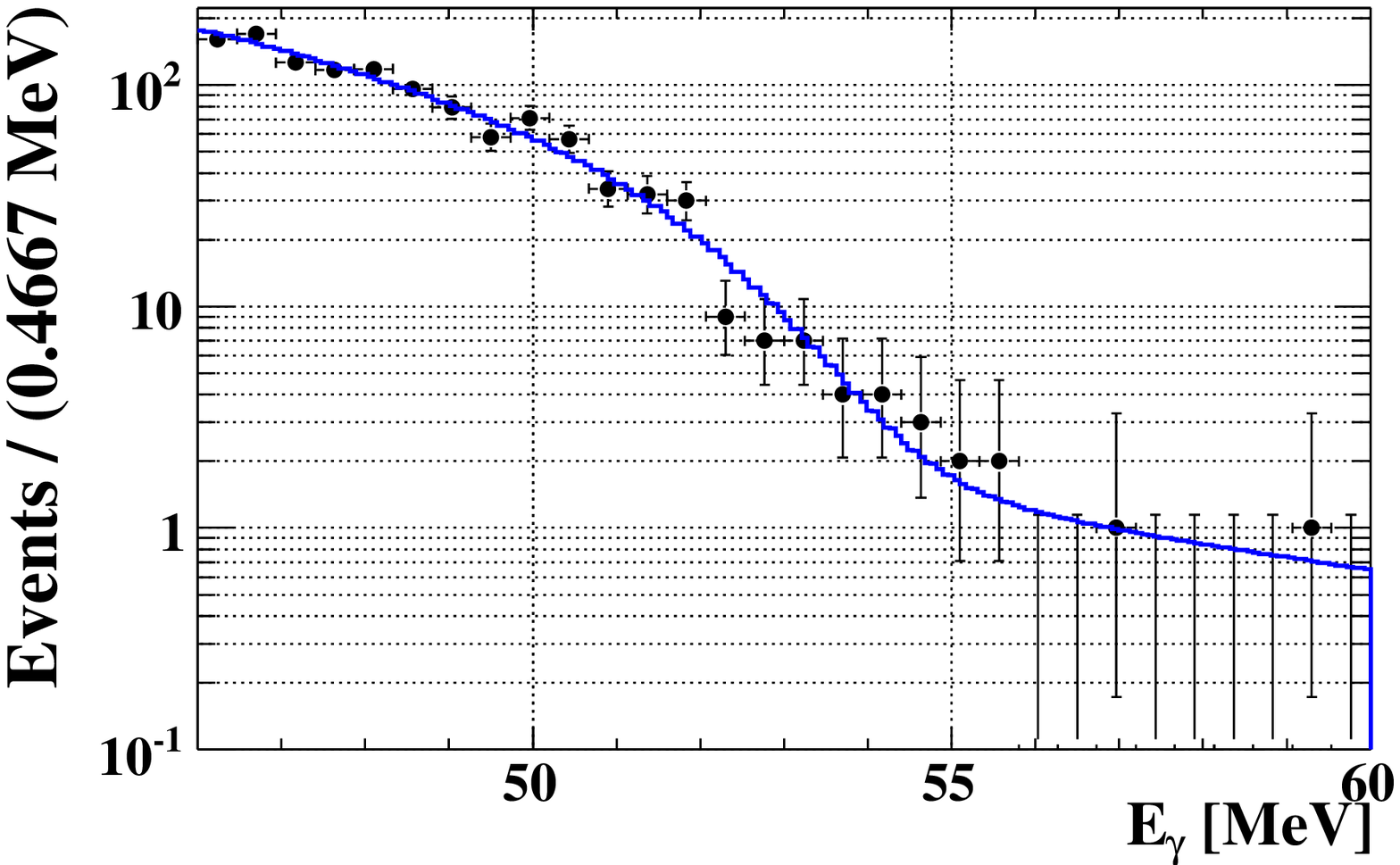}\\
\includegraphics[width=6.3cm]{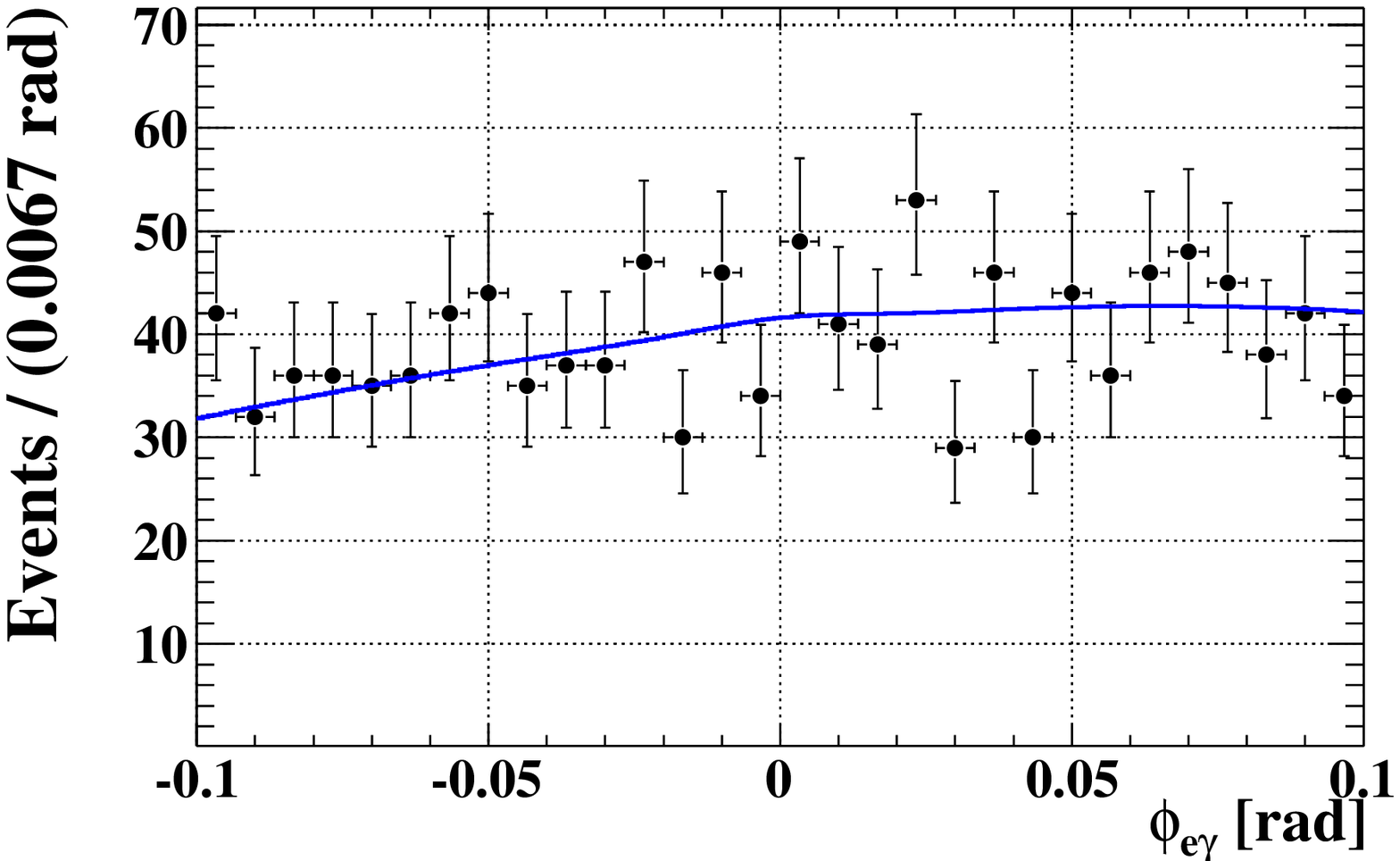}
\includegraphics[width=6.3cm]{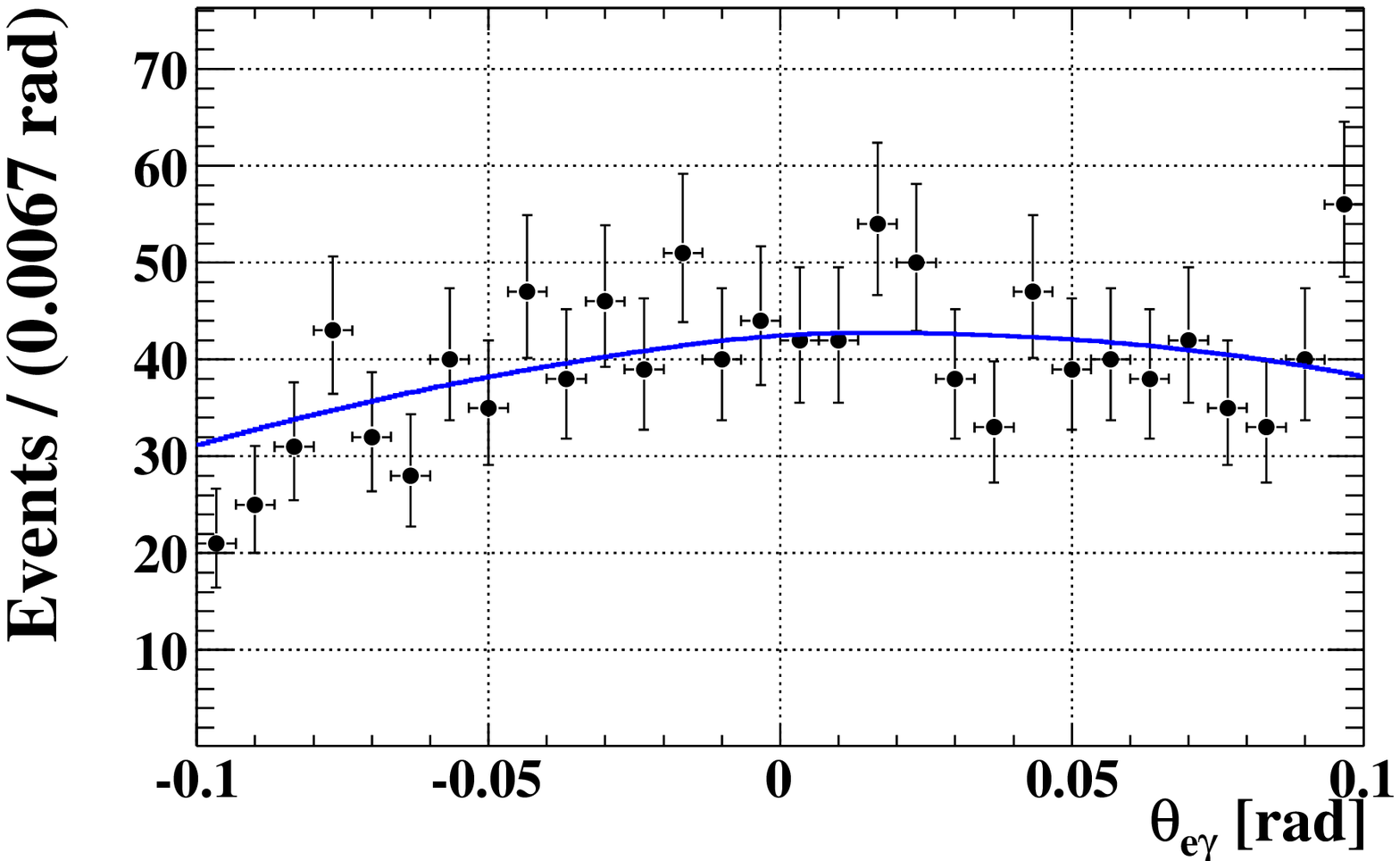}\\
\includegraphics[width=6.3cm]{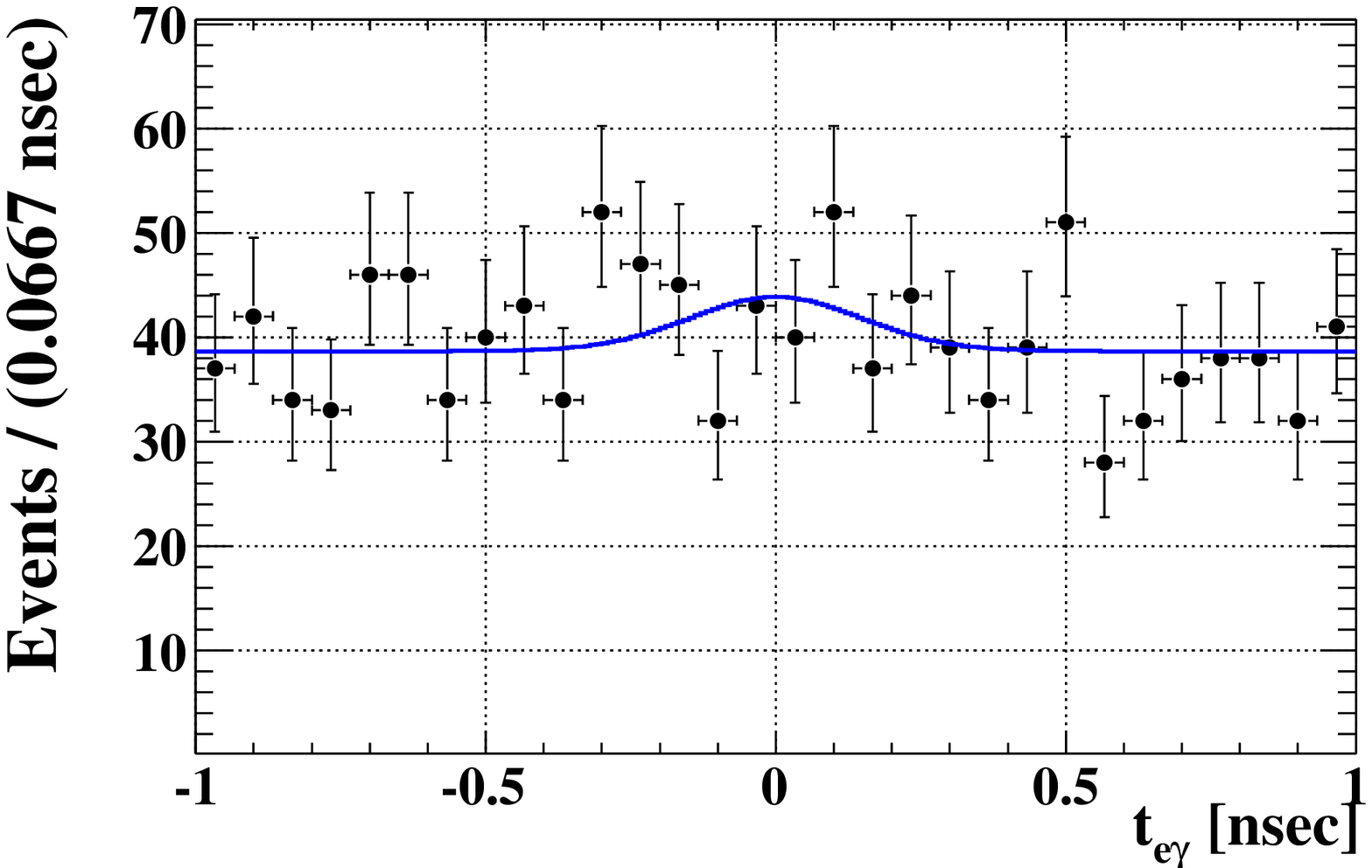}
\caption{Fit to the data in the analysis window, projected on the distributions of positron energy $E_{e}$ (top left), 
photon energy $E_{\gamma}$ (top right), positron-photon angle $\phi_{e\gamma}$ (center left), positron-photon angle
$\theta_{e\gamma}$ (center right) and positron-photon time difference $t_{e\gamma}$ (bottom).}
\label{fig:fit}
\end{center}
\end{figure}

The number of RDM, accidentals and signal events
 is determined by an extended maximuum likelihood fit to the five
variables described above in the region 46 MeV   $<E_{\gamma}<$  60 MeV, 50 MeV $< E_{e} <$
56 MeV, $|t_{e\gamma}|<$ 1 ns, $|\theta_{e\gamma}|<$ 100 mrad and $|\phi_{e\gamma}|<$  100 mrad.
The signal PDF is the product of the distributions described in~\ref{sec:res}, as they are not correlated.
For RDM, the signal PDF is taken for $t_{e\gamma}$ distribution and the other are evaluated by
folding the theoretical RDM spectrum~\cite{Kuno:1999jp} with the detector response functions extracted from
data. The PDFs for accidental events are determined on sideband data. The event distributions
of the five observables in the analysis window is shown in Fig.~\ref{fig:fit} with the fit superimposed.

The 90$\%$ confidence level intervals for the number of signal $N_{\rm{sig}}$ and RDM $N_{\rm{RDM}}$
events are determined
by means of a Feldman-Cousins approach~\cite{Feldman:1997qc}. A contour of 90 $\%$ C. L. in the $(N_{\rm{sig}},N_{\rm{RMD}})$
plane is determined with Toy MC so that, on each point of the contour, 90$\%$ of the simulated experiments
give a likelihood ratio larger than the one obtained on data. The limit on $N_{\rm{sig}}$ is extracted projecting
the contour on the $N_{\rm{sig}}$-axis and is $N_{\rm{sig}}<$14.7 at 90 $\%$ C.L., where the systematic error
is already included. The number of RDM events from the best fit is $(25^{+ 17}_{-16})$ and is consistent
with the estimation of $N_{\rm{RDM}}$ obtained in sideband data, properly scaled to the signal region 
($40 \pm 8$).

The main sources of systematic comes from the uncertainty on the selection of photon in pile-up events,
the photon energy scale and the positron energy and angular resolutions.

The UL on \mutoe is calculated by normalizing the UL on $N_{\rm{sig}}$ to the number of Michel
positron counted simultaneously with the signal and using the same analysis cut, assuming that 
$BR (\mu \to e \nu \bar{\nu}) \sim$ 1. This technique has the great advantage of being independent
of the istantaneous beam rate and is nearly insensitive to positron acceptance and efficiencies 
associated with TC or DC, as signal and Michel differ slightly only due to a small momentum
dependence. The branching fraction can in fact be written as:
\begin{eqnarray}
BR( \mu^+ \to e^+ \gamma) = \frac{N_{\rm{sig}}}{N_{e\nu\bar{\nu}}} \times \frac{f^E_{e\nu\bar{\nu}}}{P}
	\times \frac{\epsilon^{\rm{trig}}_{e\nu\bar{\nu}}}{\epsilon^{\rm{trig}}_{e\gamma}} \times 
	 \frac{A^{\rm{TC}}_{e\nu\bar{\nu}}}{A^{\rm{TC}}_{e\gamma}} \times 
	 \frac{\epsilon^{\rm{DC}}_{e\nu\bar{\nu}}}{\epsilon^{\rm{DC}}_{e\gamma}} \times 
	 \frac{1}{A^{\rm{G}}_{e\gamma}} \times \frac{1}{\epsilon_{e\gamma}}
\end{eqnarray}
where $N_{e\nu\bar{\nu}}$ = 11414 is the number of detected Michel positron in the range
50 MeV $< E_e<$ 56 MeV; P = $10^7$ is the prescale factor in the trigger used to select
Michel positrons; $f^E_{e\nu\bar{\nu}} = 0.101 \pm 0.006$ is the fraction of Michel positron
spectrum above 50 MeV; $\epsilon^{\rm{trig}}_{e\nu\bar{\nu}}/\epsilon^{\rm{trig}}_{e\gamma}
= 0.66 \pm 0.03$ is the ratio of signal to Michel trigger efficiency; 
$A^{\rm{TC}}_{e\nu\bar{\nu}}/A^{\rm{TC}}_{e\gamma}  = 1.11 \pm 0.02$ is the ratio of signal
to Michel DC-TC matching efficiency; 
$\epsilon^{\rm{DC}}_{e\nu\bar{\nu}}/\epsilon^{\rm{DC}}_{e\gamma} = 1.020 \pm 0.005$ is
the ratio of signal to Michel DC reconstruction efficiency and acceptance; 
$A^{\rm{G}}_{e\gamma} = 0.980 \pm 0.005$ is the geometrical acceptance for a signal
photon given an accepted signal positron; $\epsilon_{e\gamma} = 0.63 \pm 0.04$ is the 
efficiency of photon reconstruction and selection criteria.

The trigger efficiency ratio is different from one due to the requirement of a stringent angle
matching criteria at the trigger level. The main contributions to the photon inefficiency come
from conversions before the LXe active volume and selection criteria imposed to reject pile-up
events.

The limit on the branching ratio of \mutoe decay is therefore 
\begin{eqnarray}
BR(\mu^+ \to e^+ \gamma) \le 2.8 \times 10^{-11} \qquad \qquad \rm{(90 \% C.L.)}
\end{eqnarray}
where the systematic uncertainty on the normalization is taken into account.

The sensitivity of the experiment with this data statistic and the same number of accidental
and RDM background events, assuming null signal, calculated by means of Toy MC is
$1.3 \times 10^{-11}$, which is comparable with the best limit set by the MEGA 
experiment~\cite{Brooks:1999pu}. Given this sensitivity, the probability to obtain
an UL greater than $2.8 \times 10^{-11}$ is $\sim 5 \%$.

\section{Conclusions and Prospect}
A search for the LFV process \mutoe has been performed with a BR sensitivity of $1.3 \times 10^{-11}$,
using the data taken during the first three months of run of the MEG experiment in 2008.
With this sensitivity, which is comparable with the best UL in the world set by the MEGA
experiment, a blind likelihood analysis yields an UL on the BR of 
$BR(\mu^+ \to e^+ \gamma) \le 2.8 \times 10^{-11}$ $\rm{(90 \% C.L.)}$.

A new run has been performed between October and December 2009, where the experiment
could exploit an improved electronics with better timing and reduced noise, improved trigger 
efficiency and a smooth operation of all drift chambers, which resulted in an increase of
a factor 3-4 in efficiency. Moreover, a better understanding of the detector performances,
thanks to the experience gained from the first run, allowed to improve several other
efficiencies and resolutions. The corresponding expected sensitivity for this dataset
is ${\cal O}(10^{-12})$. A new data taking is starting in May 2010 with an stable an
improved detector and MEG is expected to continue running in 2010-2011 for the 
final ${\cal O}(10^{-13})$ goal.

\end{document}